\documentclass[aps,prd,reprint,nofootinbib,longbibliography]{revtex4-1}

\usepackage{amsmath}
\usepackage{mathtools}
\usepackage{graphicx}
\usepackage[normalem]{ulem}
\usepackage[com pat=1.1.0]{tikz-feynman}
\usepackage{tikz}
\usepackage{aas_macros}
\usetikzlibrary{external}
\usepackage{comment}
\usetikzlibrary{arrows}
\usepackage[colorlinks=true, allcolors=blue]{hyperref}

\begin{document}

\title{Where, when and why: occurrence of fast-pairwise collective neutrino oscillation in three-dimensional core-collapse supernova models}

\author{Hiroki Nagakura}
\email{hirokin@astro.princeton.edu}
\affiliation{Department of Astrophysical Sciences, Princeton University, 4 Ivy Lane, Princeton, NJ 08544, USA}
\author{Lucas Johns}
\email[]{NASA Einstein Fellow \\ ljohns@berkeley.edu}
\affiliation{Department of Physics, University of California, Berkeley, CA 94720, USA}
\author{Adam Burrows}
\affiliation{Department of Astrophysical Sciences, Princeton University, 4 Ivy Lane, Princeton, NJ 08544, USA}
\author{George M. Fuller}
\affiliation{Department of Physics, University of California, San Diego, La Jolla, California 92093, USA}

\begin{abstract}
Fast-pairwise collective neutrino oscillation represents a key uncertainty in the theory of core-collapse supernova (CCSN). Despite the potentially significant impact on CCSN dynamics, it is usually neglected in numerical models of CCSN because of the formidable technical difficulties of self-consistently incorporating this physics. In this paper, we investigate the prospects for the occurrence of fast flavor conversion by diagnosing electron neutrino lepton number (ELN) crossing in more than a dozen state-of-the-art three-dimensional CCSN models. ELN crossings is a necessary condition for triggering flavor conversion. Although only zeroth and first angular moments are available from the simulations, our new method enables us to look into the angular distributions of neutrinos in momentum space and provide accurate insight into ELN crossings. Our analysis suggests that fast flavor conversion generally occurs in the post-shock region of CCSNe, and that explosive models provide more favorable conditions for the flavor conversion than failed CCSNe. We also find that there are both common and progenitor-dependent characteristics. Classifying ELN crossings into two types, we analyze the generation mechanism of each case by scrutinizing the neutrino radiation field and matter interactions. We find key ingredients of CCSN dynamics driving the ELN crossings: proto-neutron star (PNS) convection, asymmetric neutrino emission, neutrino absorptions and scatterings. This study suggests that we need to accommodate fast flavor conversions in realistic CCSN models.
\end{abstract}
\maketitle

\section{Introduction}\label{sec:intro}
Uncovering the explosion mechanism of core-collapse supernova (CCSN) requires detailed treatment of micro- and macrophysics and non-linear feedback. Direct numerical simulations provide a means by which the complex system can be investigated in detail. Multi-dimensional neutrino-radiation hydrodynamic simulations with sophisticated input physics comprise the general ab-initio approach. Keeping in mind that numerical models of CCSN are sensitive to treatments of input physics and methodologies, we note that many state-of-the-art three-dimensional (3D) CCSN simulations now yield successful explosions, in which the combined effects of neutrino heating and fluid instabilities play crucial roles in driving runaway shock wave expansion (see, e.g., \cite{2014ApJ...786...83T,2015ApJ...807L..31L,2016ApJ...831...98R,2018ApJ...855L...3O,2019MNRAS.482..351V,2019ApJ...873...45G,2019MNRAS.484.3307M,2019MNRAS.490.4622N}). These results support the hypothesis that the delayed neutrino-heating mechanism may be the correct explosion mechanism.

However, it is premature to claim this is the answer. Almost every CCSN simulation has assumed that no neutrino flavor mixing occurs. Neglect of this physics was usually justified by pointing to the fact that the flavor-diagonal components of the neutrinos' flavor Hamiltonian are dominated by the matter potential, which reduces the effective neutrino mixing angles in the medium (see e.g., \cite{1987ApJ...322..795F,2000PhRvD..62c3007D}). Linear stability analysis of the quantum kinetic equation (QKE) for neutrino transport suggests, however, that neutrino self-interactions, which should also be taken into account in a dense neutrino medium, may trigger instabilities of flavor conversion. Once instabilities kick in, mixing would occur very rapidly on timescales much shorter than the dynamical timescale of CCSN evolution (see, e.g., \cite{2006PhRvD..74j5014D,2010ARNPS..60..569D,2016NCimR..39....1M,2020arXiv201101948T}). As a result, the energy spectrum and angular distribution of each neutrinos species may be substantially impacted. Consequently, these changes affect the neutrino-matter interactions, i.e., they might influence shock revival. It is also of importance to gauge the sensitivity of explosive nucleosynthesis and neutrino signals to these collective neutrino oscillations. Although there remain many technical challenges, these arguments provide strong motivation for tackling the problems associated with collective neutrino oscillation.

Fast-pairwise collective neutrino oscillation (fast neutrino flavor conversion), initially found by \cite{2005PhRvD..72d5003S}, has recently received increased attention (see, e.g., \cite{2017PhRvL.118b1101I,2017JCAP...02..019D,2017ApJ...839..132T,2018PhRvD..98j3001D,2019PhRvD.100d3004A,2019PhRvD..99j3011D,2019ApJ...886..139N,2020PhLB..80035088M,2020PhRvD.101b3018D,2020PhRvR...2a2046M,2020PhRvD.101d3009J,2020PhRvD.102j3017J,2020PhRvD.101d3016A,2020arXiv201101948T,2021PhRvD.103f3002S,2021PhRvD.103h3013R,2021PhRvD.103f3033A,2021PhRvD.103f3013C,2021arXiv210101226B,2021arXiv210315267M,2021arXiv210312743S,2021arXiv210615622S}). As the name suggests, it is the fastest growing mode among the flavor field instabilities in a dense neutrino gas. The growth rate is roughly proportional to the number density of the electron neutrino lepton number (ELN) in the CCSN environment. According to previous studies, the necessary condition triggering the instability is ELN angular crossings \cite{2021arXiv210315267M}, in which the sign of the ELN varies with angle in momentum space. Hence, searching for ELN crossings in the neutrino data of CCSN simulations is important.

An ELN crossing search requires detailed information on the flavor-dependent neutrino angular distributions, which are only available from CCSN models employing full Boltzmann neutrino transport. Although such demonstrations with full Boltzmann neutrino transport have recently appeared in the literature \cite{2017ApJ...839..132T,2019PhRvD.100d3004A,2019PhRvD..99j3011D,2019ApJ...886..139N,2020PhRvD.101b3018D,2020PhRvD.101d3016A}, most 3D CCSN simulations employ approximate neutrino transport to reduce the computational cost. For instance, two-moment methods with analytical closure relations are a frequently employed approach. Although this approximation enables us to carry out long-term 3D simulations for multiple progenitors, information on the full angular distributions of neutrinos is abandoned; instead, a few low angular moments are obtained. The absence of full angular information precludes carrying out ELN crossing searches, indicating that alternative approaches are indispensable for the study of fast flavor conversion in these numerical CCSN models.

During the last few years, some alternative methods have been proposed in the literature. The simplest approach is the {\it zero mode search} \cite{2018PhRvD..98j3001D}. This method takes advantage of the characteristics of fast flavor conversion; the stability of the $k=0$ mode ($k$ denotes the wave number in the corotating frame) can be determined by the zeroth, first, and second angular moments. Another approach has been proposed in \cite{2020JCAP...05..027A}. In this method, the stability of fast flavor conversion is determined by angular integrated quantities of neutrinos, in which the integration is carried out with polynomial functions. This allows one to determine the existence of ELN crossings from a finite number of angular moments.

More recent studies in \cite{2021PhRvD.103l3012J,2021arXiv210405729N} have suggested, however, that these methods have several weaknesses due to the limited information from angular moments. In essence, these approaches are not capable of providing accurate insights into the stability of flavor conversions in the region $\kappa \gtrsim 0.5$, where $\kappa$ denotes the flux factor of neutrinos. The limitation is due to the fact that the angular structures of neutrinos in forward-peaked distributions are characterized mainly by high angular moments. The study of \cite{2021arXiv210405729N} further revealed that the angular distributions of incoming neutrinos are insensitive to the lower moments. This is an intrinsic limitation that inevitably plague any approach based on moment methods, resulting in the reduction in the accuracy of the ELN crossing search.

These facts motivated us to develop a better alternative method by which to search for ELN crossings \cite{2021arXiv210602650N}. There are four important elements in the method. First, the angular distribution of neutrinos is constructed from the zeroth and first angular moments by using a method calibrated by a full Boltzmann neutrino transport code \cite{2014ApJS..214...16N,2017ApJS..229...42N,2019ApJ...878..160N,2021arXiv210405729N}. Second, we use a ray-tracing method to compensate for the limitation of moment-based approach. This hybrid approach is capable of providing more accurate angular information on neutrinos than moments alone, and this facilitates our understanding of the origin of ELN crossings. Third, our ELN crossing searches are carried out using the distribution functions of neutrinos, implying that the energy-dependent features of neutrino angular distributions are encoded in the search procedure. This distinct from other methods \cite{2018PhRvD..98j3001D,2020JCAP...05..027A}, in which only energy-integrated quantities are used. Fourth, our ELN crossing search is computationally much cheaper than full Boltzmann neutrino transport. This enables us to apply our method to many 3D CCSN models. We will describe our method in Sec.~\ref{sec:modelandmethod}.

In this paper, we present a detailed study of ELN crossings for state-of-the-art 3D CCSN simulations with two-moment neutrino transport. We cover more than a dozen of CCSN models, that include different progenitor masses, different (angular) resolutions, and exploding/non-exploding cases. We report both common and progenitor-dependent trends of the appearance of ELN crossings among these models. As we show in Sec.~\ref{sec:result}, our results vary from those obtained in previous studies \cite{2021PhRvD.103f3033A}. We validate our procedure by providing clear physical interpretations of results. Although there are some limitations of both our CCSN simulations and ELN crossing search method, the present study suggests that fast flavor conversions commonly occur in the post-shock regions of CCSNe. This result motivates more detailed studies of fast flavor conversion.

This paper is organized as follows. We summarize our 3D CCSN models in Sec.~\ref{subsec:3DCCSN}, and also provide a brief overview of our ELN crossing search method in Sec.~\ref{subsec:ELNcsearch}. All results and analyses regarding ELN crossing searches are encapsulated in Sec.~\ref{sec:result}. In that section, we first report the overall properties of ELN crossings (Sec.~\ref{subsec:overall}), and then carry out an in-depth analysis of the generation mechanism of each type of ELN crossing (Sec.~\ref{subsec:mechanism}). Finally, we summarize our conclusions in Sec.~\ref{sec:summary}.

\section{Models and methods}\label{sec:modelandmethod}

\begin{figure}
    \includegraphics[width=\linewidth]{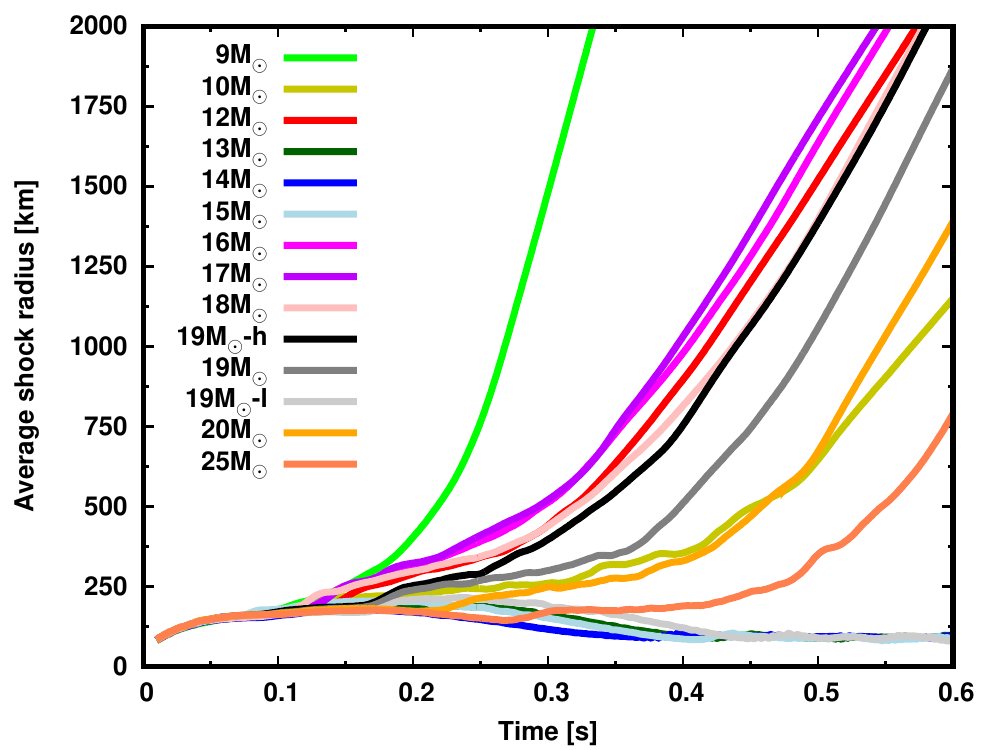}
    \caption{Evolution of the angle-averaged shock wave as a function of time for 3D CCSN models. The time is measured from core bounce. Color distinguishes the CCSN model.}
    \label{graph_timetrajectories_shockradii_FFC}
\end{figure}

\begin{figure}
    \includegraphics[width=\linewidth]{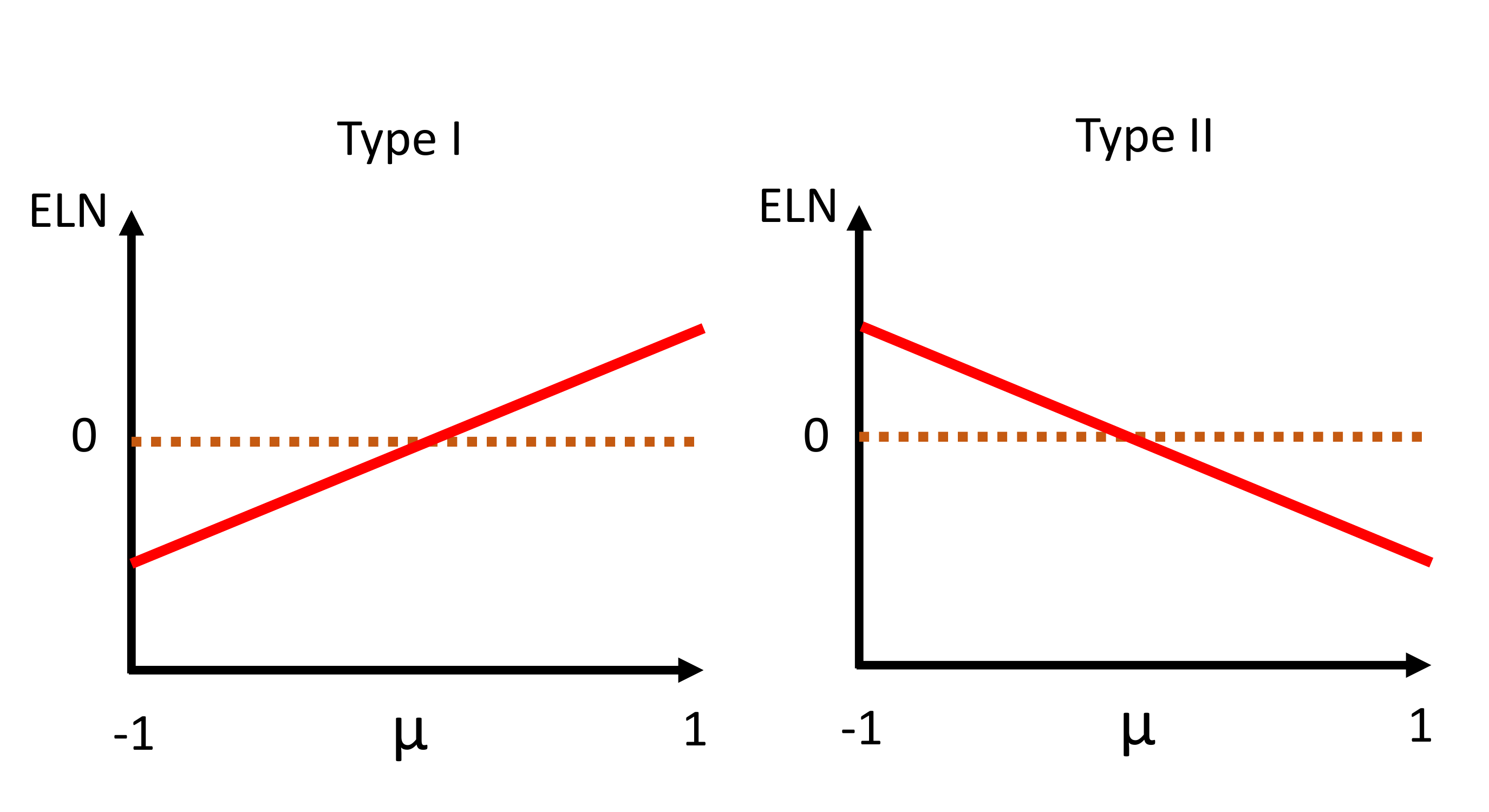}
    \caption{Classifying the type of ELN crossings. The horizontal axis denotes the directional cosine of the neutrino propagation measured from the radial basis. In the Type-I crossing (left panel), $\nu_e$ dominates over $\bar{\nu}_e$ at $\mu=1$ whereas $\bar{\nu}_e$ overwhelms $\nu_e$ at $\mu=-1$. The Type-II crossing (right panel) corresponds to the opposite case.}
    \label{graphType}
\end{figure}

\subsection{3D CCSN models}\label{subsec:3DCCSN}
In this study, we employ fourteen 3D CCSN models simulated by a state-of-the-art multi-D neutrino-radiation-hydrodynamic code: F{\sc{ornax}} \cite{2019ApJS..241....7S}. The neutrino transport is solved under a multi-energy two-moment method, in which the modern neutrino-matter interactions are incorporated by following \cite{2006NuPhA.777..356B,2017PhRvC..95b5801H}. The energy-dependent zeroth and first angular moments are the basic outputs describing the neutrino fields our CCSN simulations. Our ELN crossing search procedure is effected by post processing this data.

The majority of our CCSN models exhibit successful explosions (see Fig.~\ref{graph_timetrajectories_shockradii_FFC}). The dynamics of these models  is essentially in line with the delayed neutrino-heating mechanism aided by multi-D fluid instabilities. We also consider four non-exploding models: these correspond to $13, 14, 15$ solar mass progenitors and a $19$ solar mass case simulated with a low angular resolution. We refer readers to our previous papers (e.g., \cite{2020MNRAS.491.2715B,2019MNRAS.490.4622N}) for details of the progenitor- and resolution-dependence of the simulation results. Although these non-explosion outcomes are not definitive, some characteristic features of failed CCSNe would be imprinted in these models. As we discuss in Sec.~\ref{subsec:overall}, shock revivals would be an important factor to facilitate ELN crossings. Although the trend is consistent with that reported from \cite{2021PhRvD.103f3033A}, our analysis calls into question their interpretation (see See Sec.~\ref{subsec:mechanism} for more detail). In the next section, we will reveal the role of multi-D effects such as convection and asymmetric neutrino emission in the appearances of ELN crossings. Providing physical insights into the connection between CCSN dynamics and ELN crossings is the main purpose of this paper.

\subsection{ELN crossing search}\label{subsec:ELNcsearch}

Here, we briefly describe the essence of our ELN crossing search. We refer readers to \cite{2021arXiv210602650N} for the details. We start by using the zeroth and first angular moments of neutrinos obtained from CCSN simulations to compute the energy-dependent flux factor ($\kappa$). The flux factor is one of the key quantities in our method. By using it, we construct angular distributions of neutrinos. Although the finite number of moments, in general, preclude full reconstruction of angular distributions, adequately detailed distributions can be obtained by using the reconstruction method developed in \cite{2021arXiv210405729N}. The success of this method can be attributed to the fact that the neutrino radiation field in CCSNe has specific characteristics that relate $\kappa$ to higher angular moments. In \cite{2021arXiv210405729N}, we developed a fitting formula with which to determine the shape of angular distributions of neutrinos from $\kappa$; the coefficients are chosen so as to minimize the deviations from those obtained by full Boltzmann neutrino transport in \cite{2017ApJ...847..133R}\footnote{The best fit parameters are publicly available: \url{https://www.astro.princeton.edu/~hirokin/scripts/data.html}}. We confirmed that our method is capable of providing reasonable angular distributions of neutrinos even in cases with multi-D CCSN models\footnote{There is a caveat, however. The fitting formula provided in our method \cite{2021arXiv210405729N} corresponds to the azimuthal-averaged angular distribution of neutrinos, i.e., the non-axisymmetric structures are abundant. On the other hand, in our previous study \cite{2019ApJ...886..139N}, we found that ELN crossings occur in the non-axisymmetric direction. Although this shortage can be overcome, we do not improve it. See text for details.}. This was validated by making detailed comparisons to the results of axisymmetric CCSN simulations with full Boltzmann neutrino transport \cite{2019ApJ...880L..28N}. We refer readers to \cite{2021arXiv210405729N} for details of this validation.

As we have already pointed out, however, angular distributions of incoming neutrinos in optically thin regions are not well reconstructed. Consequently, we do not adopt the reconstructed distributions; instead, they are complemented with a ray-tracing method. Although ray-tracing neutrino transport is, in general, very computationally expensive, we solve them only along the radial ray in the inward direction (at $\mu=-1$, where $\mu$ denotes the cosines of the angles measured from the radial basis), which substantially reduces the computational cost. The neutrino-matter interaction rates are computed by using the fluid data taken from the CCSN simulations.

The ray-tracing computation provides the neutrino distribution function for $\mu=-1$ ($f_{\rm in}$)\footnote{It should be mentioned that the ELN at $\mu=-1$ at optically thick region can be computed accurately from the reconstructed angular distributions computed from zeroth and first angular moments; hence, we adopt them in the region. For the intermediate region (semi-transparent region), we determine the ELN by mixing the results of reconstructed angular distribution and those obtained from the ray-tracing method. See \cite{2021arXiv210602650N} for details.}. The ELN at $\mu=-1$ is obtained by carrying out the energy-integration of $f_{\rm in}$ for electron-type neutrinos ($\nu_e$) and their anti-particles ($\bar{\nu}_e$). On the other hand, the ELN in the direction of $\mu=1$ is computed from the reconstructed angular distributions at $\mu=1$ ($f_{\rm out}$). It should be mentioned that $\kappa$ has a strong correlation to $f_{\rm out}$ (see \cite{2021arXiv210405729N}), which allows us to compute the ELN at $\mu=1$ accurately solely from moments.

We determine the appearance of ELN crossings by comparing the sign of the ELN at $\mu=-1$ and $1$. Opposite signs signal appearances of ELN crossings. One may wonder if this criterion misses cases with an even number of crossings, since the sign of the ELN at $\mu=-1$ and $1$ should be the same in these cases. However, we have witnessed in CCSN simulations with full Boltzmann neutrino transport that the majority of ELN crossings appearing in CCSNe are single crossings \cite{2019ApJ...886..139N}. Although our method, in principle, is capable of handling multiple ELN crossings (and non-axisymmetric crossings) by increasing the number of rays in the ray-tracing method, this requires as much computational cost as that of multi angle neutrino transport. Hence, we do no make these improvements in our present ELN crossing search. We leave more quantitative expositions of this physics to CCSN simulations with full Boltzmann neutrino transport.

A few points must be made regarding the present study. Since a consensus has been reached that fast conversion commonly occurs in the pre-shock region at $\gtrsim 100$~ms after core bounce \cite{2020PhRvR...2a2046M}, we only focus on the post-shock region in this paper. It should also be mentioned that we analyze the neutrino radiation fields in the region of $R < 1,000$ km, and we set the upper limit of $\kappa$ as 0.99 in this study. This limitation is due to the fact that there are some numerical artifacts in the output of Fornax simulations at $R \gtrsim 1,000$ km or in the region of large $\kappa$ ($\gtrsim 0.99$), that leads to unphysical outcomes in our ELN crossing analysis. These limitations, however, do not compromise our capture of the qualitative trend of ELN crossings in 3D CCSN models.

Let us close this section by classifying ELN crossings into two types (see Fig.~\ref{graphType}). This classification is useful for discussing relation of ELN crossings to CCSN dynamics (see Sec.~\ref{sec:result} for more detail). In the following analysis, we assume that all ELN crossings are single crossings\footnote{There is a caveat, however. Our method may detect the odd-number of multiple crossings, since the sign of ELN at $\mu=-1$ and $1$ is opposite in these cases. It should be stressed that such detailed structures are unclear without full Boltzmann neutrino transport; hence, we assume the single crossing in this study. On the other hand, the multiple crossings are subdominant in CCSN environment, as mentioned already. Our assumption is, hence, reasonable to understand the characteristics of ELN-crossings qualitatively.}. With this assumption there are two types of crossings: one of them is that the ELN at $\mu=1$ (outgoing) is positive, i.e., $\nu_e$ dominates over $\bar{\nu}_e$; while $\bar{\nu}_e$ overwhelms $\nu_e$ at $\mu=-1$, i.e., ELN is negative there. We call this case a {\it Type I crossing}. The other type, here deemed a {\it Type II crossing}, is just the inverse of the Type I case.

\section{Results}\label{sec:result}
\subsection{Overall properties}\label{subsec:overall}

\begin{figure*}
  \begin{minipage}{0.9\hsize}
    \includegraphics[width=\linewidth]{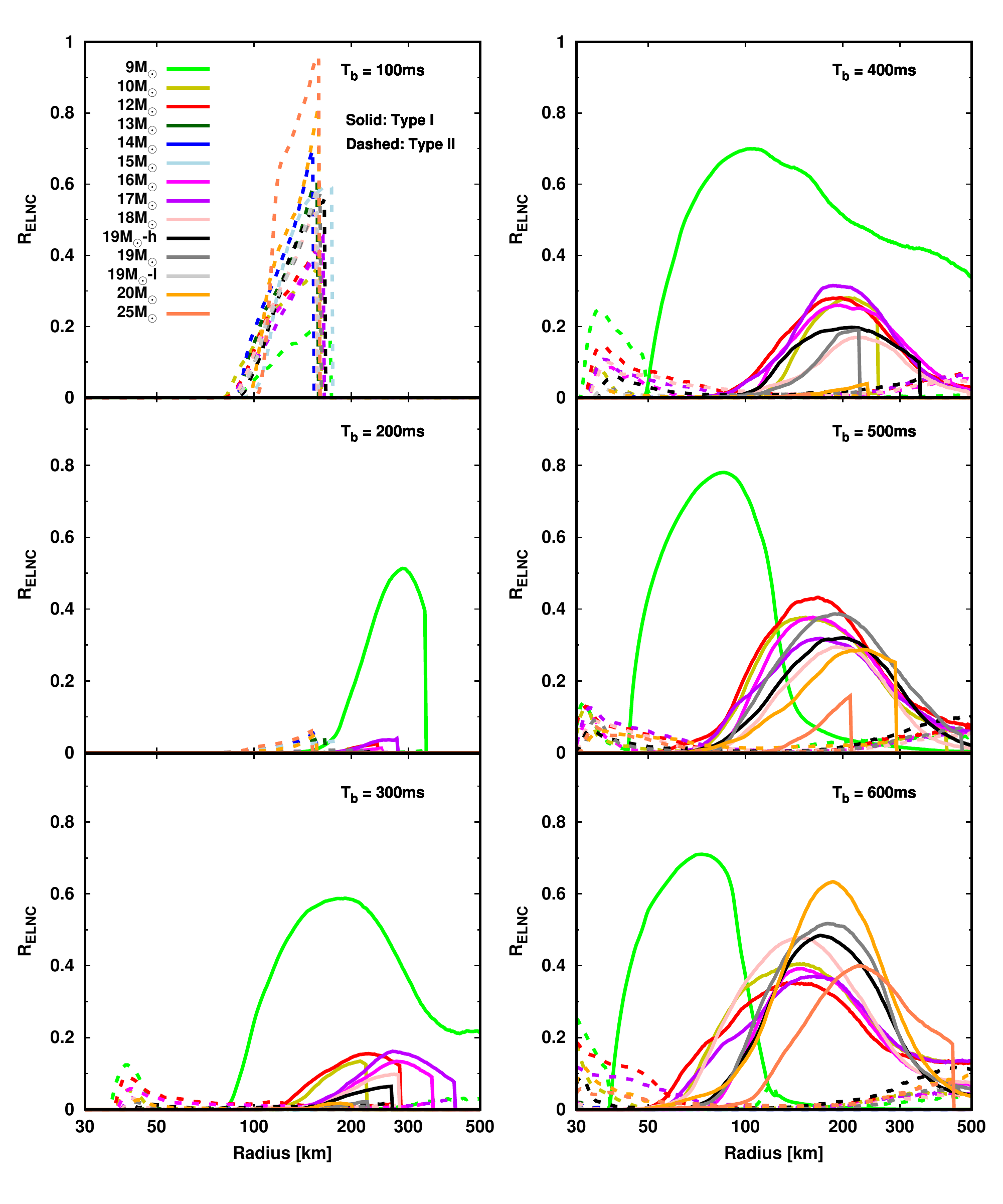}
    \caption{Radial profiles of $R_{\rm ELNC}$ (see also Eq.~\ref{eq:ELNrate}). The color distinguishes CCSN models, and the color-code is the same as that in Fig.~\ref{graph_timetrajectories_shockradii_FFC}. The line-type represents the type of ELN crossings; the solid and dashed lines correspond to Type-I and Type-II crossings, respectively. We selected 6 different time snapshots, which run from $100$ ms to $600$ ms every $100$ ms from top left to bottom right. To highlight the post-shock region, we mask the pre-shock region by setting $R_{\rm ELNC} = 0$ in this figure.}
    \label{graph_Angave_rvsELNC}
  \end{minipage}
\end{figure*}

\begin{figure*}
  \begin{minipage}{0.8\hsize}
    \includegraphics[width=\linewidth]{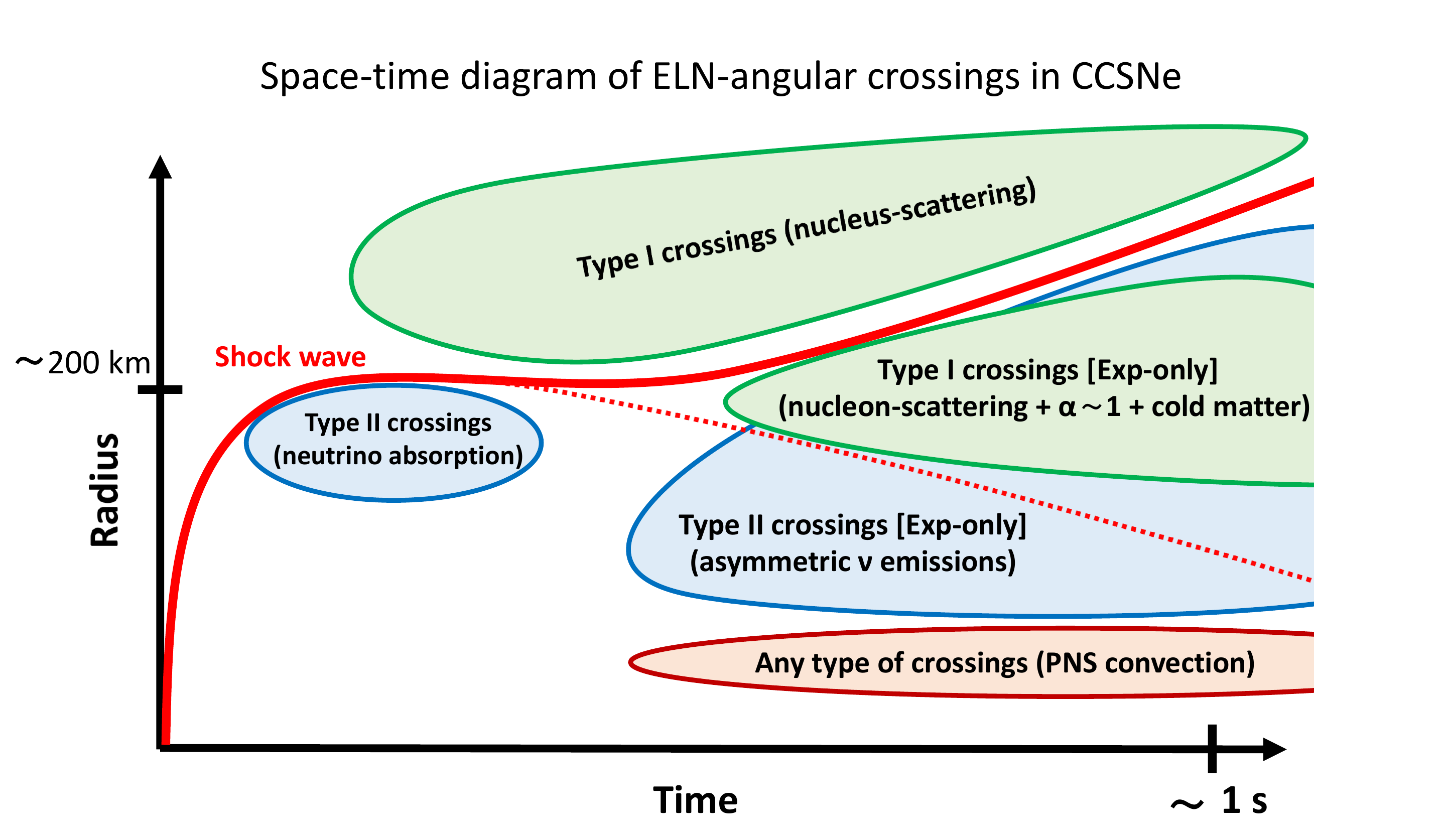}
    \caption{Space-time diagram for appearance of ELN crossings. The bold red line portrays a time trajectory for the shock wave in exploding models. The thin and dashed line represents the counterpart of shock trajectory for non-exploding models. The color code for enclosed regions distinguishes types of ELN crossing. The green, blue, and brown color denote Type I, Type II, and any type of crossings, respectively. In each region, we provide some representative characteristics of ELN-crossings. The remark "Exp-only" denotes that the ELN-crossing appears only in exploding models. See text for more detail.}
    \label{PhaseDiagram}
  \end{minipage}
\end{figure*}

Figure~\ref{graph_Angave_rvsELNC} illustrates the overall trend of ELN crossings appearing in our CCSN models. It displays the radial profile of $R_{\rm ELNC}$ which represents the occupation rate of each type of ELN crossings with respect to the entire $4 \pi$ sphere. More precisely, $R_{\rm ELNC}$ is defined at each radius as
\begin{equation}
R_{\rm ELNC} \equiv \frac{1}{4 \pi} \int d \Omega N_{i}(\Omega),
\label{eq:ELNrate}
\end{equation}
where $\Omega$ denotes the solid angle; $N_{i}$ (the subscript "$i$" is either Type I or II, see also Fig.~\ref{graphType}) is 1, if the crossing appears, otherwise it is zero.

We find that Type-II ELN crossing appears at the early post-bounce phase ($\sim 100$~ms after bounce); see the top left panel of Fig.~\ref{graph_Angave_rvsELNC}. Interestingly, this crossing appears in all CCSN models, and is
 commonly located right behind the shock wave. The disparity of neutrino absorptions between $\nu_e$ and $\bar{\nu}_e$ is responsible for driving the crossings. The high mass accretion rate at this phase also helps facilitate the generation of crossings. See Sec.~\ref{subsec:mechanism} for more detail.

The Type-II crossings gradually disappear with time; instead, Type-I crossings emerge in some models (see the panels at $\geq 200$ ms in Fig.~\ref{graph_Angave_rvsELNC}). We find that the crossing is closely associated with shock revival; indeed, it does not appear in non-explosion models. We also find that the Type-I crossing emerges behind the shock wave once the shock wave enters into a runaway expansion phase. This trend can be clearly seen by comparing to the time evolution of the shock wave (see Fig.~\ref{graph_Angave_rvsELNC}). The ELN crossing in the 9 solar mass model stands out at $\sim 200$ ms (see the left middle panel of Fig.~\ref{graph_Angave_rvsELNC}). Shock revival already has been achieved by that time only for this model (see Fig.~\ref{graph_timetrajectories_shockradii_FFC}). Interestingly, the spatial distributions of the crossings evolve with time. As shown in Fig.~\ref{graph_Angave_rvsELNC}, the crossing is observed in the vicinity of the shock wave soon after the onset of shock revival, but it sinks into the inner region at later times. We also find that the $R_{\rm ELNC}$ has a peaked profile, i.e., the crossing disappears in the very inner regions.

It is interesting to note that Type-II ELN crossings still can be observed in some models at $\gtrsim 200$ ms. Most of the crossings tend to appear in the very inner region where Type-I crossings completely disappear (see the right panels of Fig.~\ref{graph_Angave_rvsELNC}). As such, the two types of crossings have distinct properties, indicating that the generation mechanism  for these is different. In Sec.~\ref{subsec:mechanism}, we conduct an in-depth analysis of their physical origin.

We provide a schematic space-time diagram of ELN crossings in Fig.~\ref{PhaseDiagram}. This figure summarizes the overall trends of crossings observed in our CCSN models. We note that crossings relevant to PNS convection and the pre-shock region drawn in Fig.~\ref{PhaseDiagram} are not included in Fig.~\ref{graph_Angave_rvsELNC}. There is a technical reason why we do not include the case with PNS convection in Fig.~\ref{graph_Angave_rvsELNC}. This issue will be discussed later. To facilitate the readers' understanding, the color in Fig.~\ref{PhaseDiagram} distinguishes types of ELN-crossings. Below, we turn our attention to the physical origin of ELN crossing generation.

\subsection{Generation mechanism of ELN crossings}\label{subsec:mechanism}

\begin{figure*}
  \begin{minipage}{0.8\hsize}
    \includegraphics[width=\linewidth]{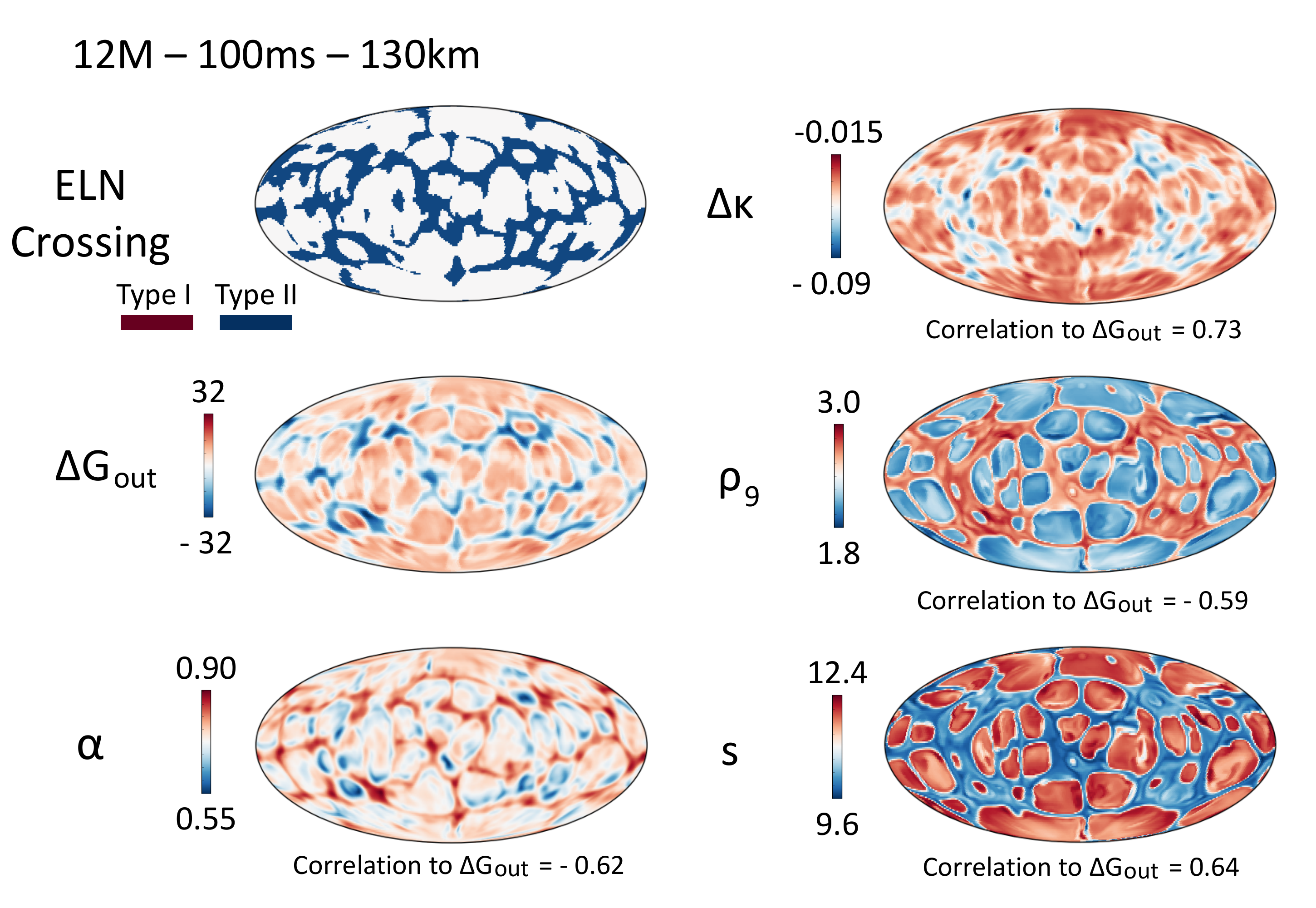}
    \caption{Mollweide projection of some important quantities characterizing ELN crossings in the12 solar mass model. We present projection maps at a radius of $130$ km for a time snapshot at $100$ ms after the bounce. The plots in the left column are ELN crossings, ELN at $\mu=1$ ($\Delta G_{\rm out}$, see Eqs.~\ref{eq:def_GoutGin}~and~\ref{eq:def_deltaGout}), and the ratio of number density of $\bar{\nu}_e$ to $\nu_e$ ($\alpha = n_{\bar{\nu}_e}/n_{\nu_e}$) from top to bottom. In the right one, we display the result of the difference of (energy-integrated) flux factor between $\nu_e$ and $\bar{\nu}_e$ ($\Delta \kappa$), baryon mass density ($\rho$), and entropy per baryon ($s$), respectively. $\rho_9$ denotes the normalized baryon mass density by $10^{9} {\rm g/cm^3}$. The values of correlation function to $\Delta G_{\rm out}$ (see Eqs.~\ref{eq:Cordef}~and~\ref{eq:CordCompodef}) are also inserted at the bottom of each Mollweide projection.}
    \label{Mol_12M_100ms_130km}
  \end{minipage}
\end{figure*}

\begin{figure}
  \begin{minipage}{0.8\hsize}
    \includegraphics[width=\linewidth]{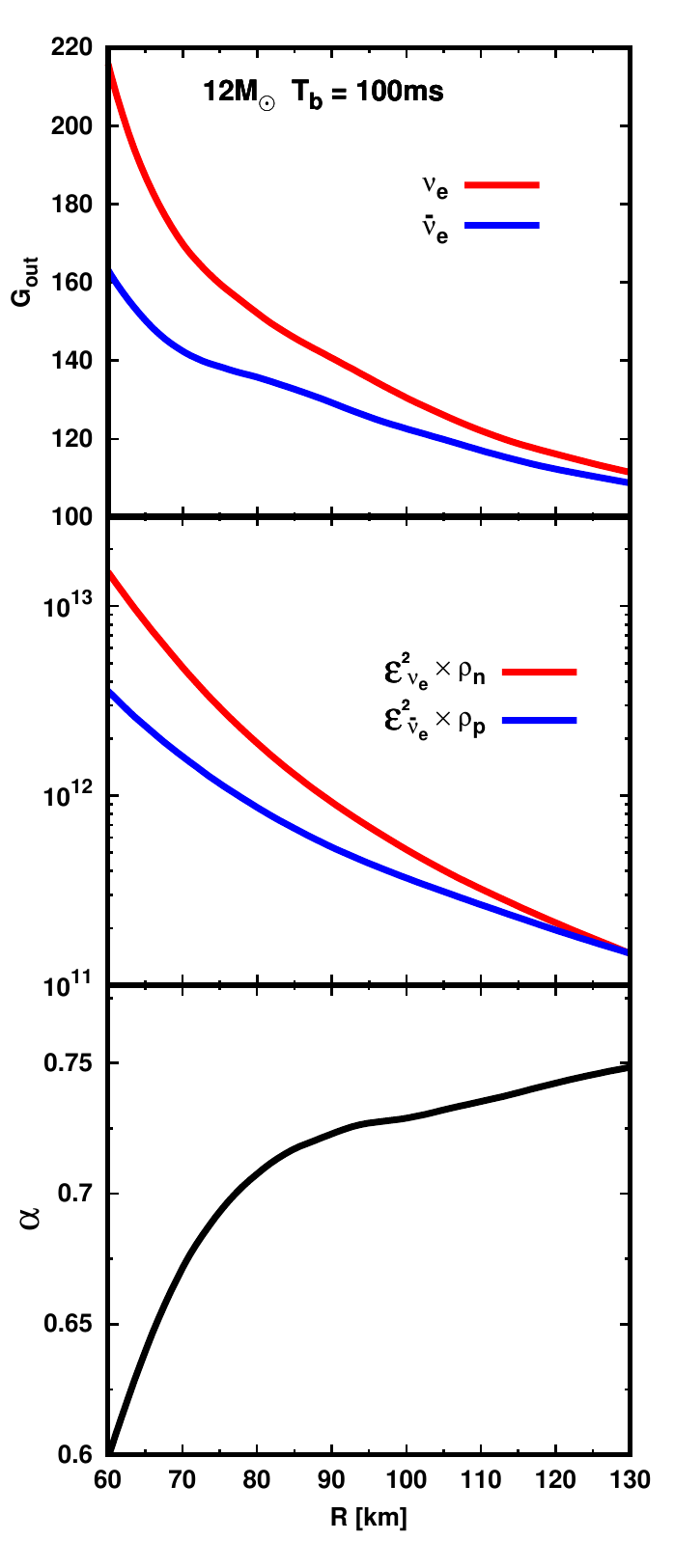}
    \caption{Radial profiles of angle-averaged quantities. In the top panel, we display $G_{\rm out}$ for $\nu_e$ (red) and $\bar{\nu}_e$ (blue). In the middle one, we show $\varepsilon^2_{\nu_e} \rho_n$ (red) and $\varepsilon^2_{\bar{\nu}_e} \rho_p$ (blue), where $\varepsilon$ denotes the average energy of each species of neutrinos. The former and the latter are roughly proportional to $\nu_e$ and $\bar{\nu}_e$ absorption rate, respectively. In the bottom panel, we display the radial profile for $\alpha$.}
    \label{graph_12M100msAngave}
  \end{minipage}
\end{figure}

\begin{figure*}
  \begin{minipage}{0.8\hsize}
    \includegraphics[width=\linewidth]{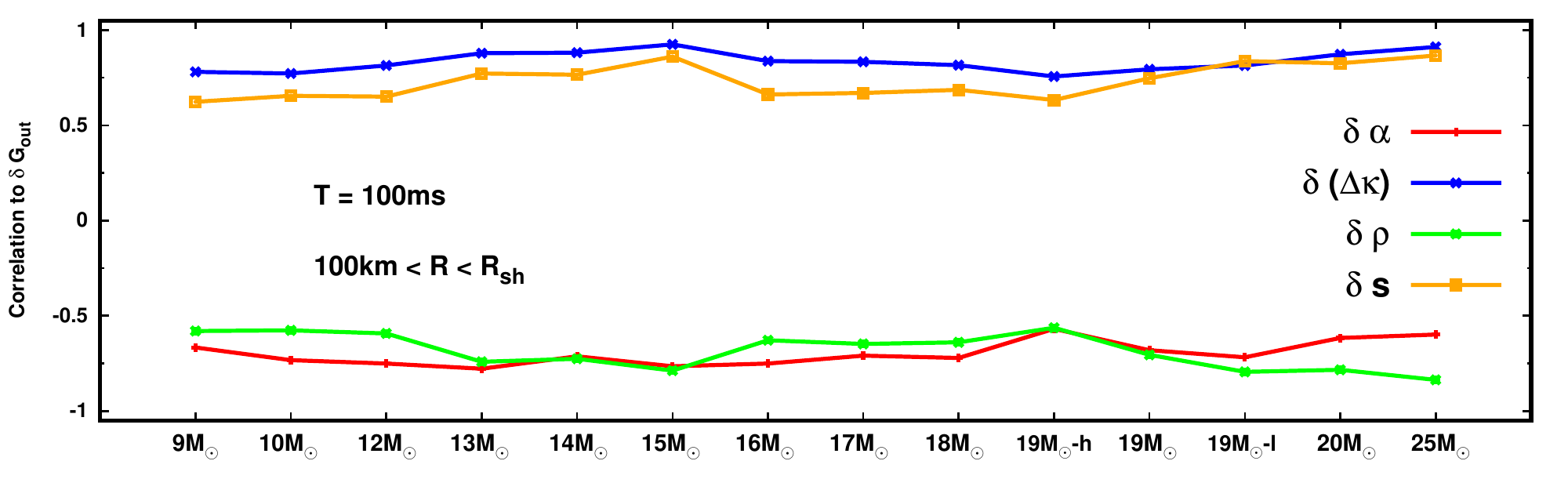}
    \caption{The value of correlation function of each quantity ($\alpha$, $\Delta \kappa$, $\rho$, and $s$) to $G_{\rm out}$. The correlation function is computed by subtracting the angle-averaged quantity ("$\delta$" denotes the deviation from the angle-averaged one, see Eq.~\ref{eq:Cordef} for more detail). We adopt the maximum (minimum) value of correlation function for the positive (negative) sign in the region of $100 {\rm km} < R < R_{\rm sh}$. The horizontal axis distinguishes the progenitor dependent.}
    \label{graph_Correlation_ELNout_100ms_100km_Rsh_prodepe}
  \end{minipage}
\end{figure*}

\begin{figure}
  \begin{minipage}{1.0\hsize}
    \includegraphics[width=\linewidth]{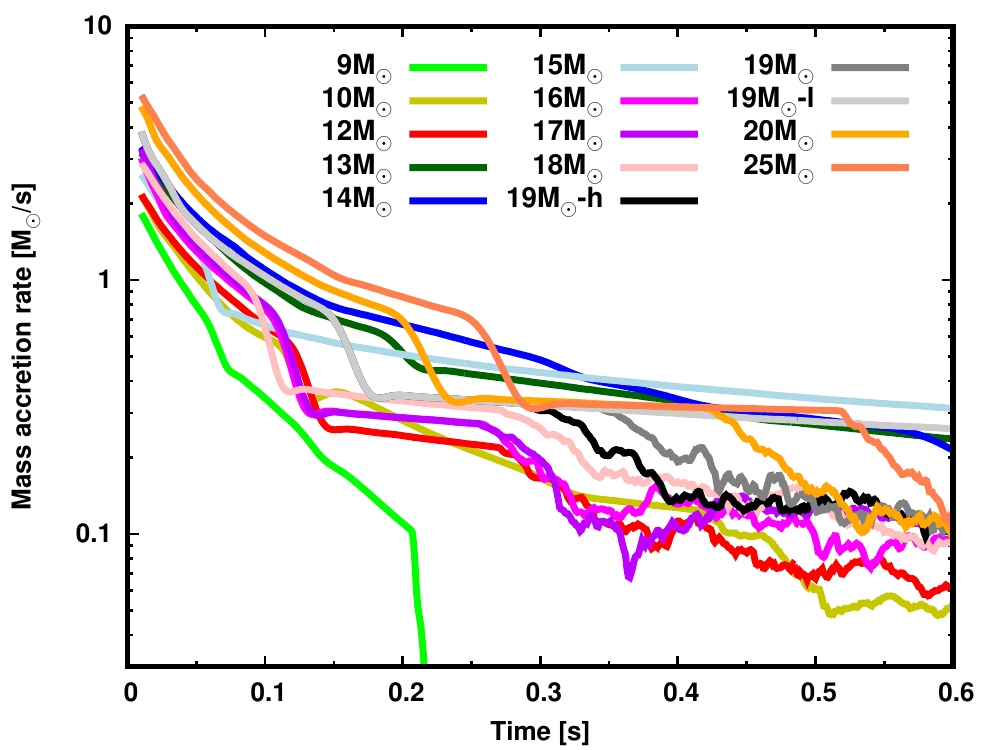}
    \caption{Time evolution of mass accretion rate that is measured at the radius of $500$ km.}
    \label{graph_Massaccretion_ELN}
  \end{minipage}
\end{figure}

\subsubsection{Type-II crossings at early post-bounce phase}\label{subsubsec:TypeIIearly}
Let us start by analyzing the Type-II crossings that appear at the early post-bounce phase ($\sim 100$ ms) in all CCSN models (see the top left panel in Fig.~\ref{graph_Angave_rvsELNC}). We first present the result from the 12 solar mass model as a representative case. The progenitor-dependence is discussed later. In Fig.~\ref{Mol_12M_100ms_130km}, we show Mollweide projections of the ELN crossing and some important quantities at $130$ km for the 12 solar mass model case. We find that the Type II crossing has a rather scattered distribution (see the top left panel). To see the trend more quantitatively, we show $\Delta G_{\rm out}$ in the left middle panel in Fig.~\ref{Mol_12M_100ms_130km}, which corresponds to the ELN at $\mu=1$. Here $\Delta G_{\rm out}$ and $\Delta G_{\rm in}$ are defined as follows. The energy-integrated number of neutrinos at $\mu=1$ and $-1$ are written as
\begin{equation}
\begin{aligned}
&G_{\rm out} = \int d (\frac{\varepsilon^3}{3}) f_{\rm out}(\varepsilon), \\
&G_{\rm in} = \int d (\frac{\varepsilon^3}{3}) f_{\rm in}(\varepsilon), \\
\end{aligned}
\label{eq:def_GoutGin}
\end{equation}
respectively, where $\varepsilon$ denotes the neutrino energy in units of MeV. We stress that both $f_{\rm out}$ and $f_{\rm in}$ in Eq.~\ref{eq:def_GoutGin} are the basic output of our angular reconstruction computation complemented by the ray-tracing method (see Sec.~\ref{subsec:ELNcsearch}). Here $\Delta G$ is the difference of the $\nu_e$ and $\bar{\nu}_e$ $G$ values:
\begin{equation}
\Delta G = G_{\nu_e} - G_{\bar{\nu}_e},
\label{eq:def_deltaGout}
\end{equation}
where we omit the subscript "out" or "in" in Eq.~\ref{eq:def_deltaGout}. As shown in Fig.~\ref{Mol_12M_100ms_130km}, we find that $\bar{\nu}_e$ dominates over $\nu_e$ in some regions (blue-colored area), and these regions are in one-to-one correspondence to the regions of Type-II crossings. The one-to-one correspondence is attributed to the fact that $\nu_e$ always overwhelms $\bar{\nu}_e$ in $\mu=-1$ (incoming) direction.

We find some interesting correlations between the Type-II crossings and other physical quantities. These correlations provide useful insight for studying the physical origin of the crossings. To quantify the correlations, we define the correlation function for $\Delta G_{\rm out}$ by following \cite{2017ApJ...851...62K,2019MNRAS.489.2227V,2021MNRAS.tmp.1539N},
\begin{equation}
{\rm X} = \frac{ {\rm Y}_{A G}  }{ {\rm Y}_{A}  \times {\rm Y}_{G}  }, \label{eq:Cordef}
\end{equation}
where
\begin{equation}
\begin{aligned}
&{\rm Y}_{A G}  = \int  d \Omega \hspace{0.5mm} \delta A (\Omega) \hspace{0.5mm} \delta (\Delta G_{\rm out} (\Omega)), \nonumber \\
&{\rm Y}_{A} = \sqrt{ \int  d \Omega \hspace{0.5mm}  \left(\delta A(\Omega) \right)^2 }, \nonumber \\
&{\rm Y}_{G}  = \sqrt{ \int  d \Omega \hspace{0.5mm} \left(\delta (\Delta G_{\rm out}(\Omega)) \right)^2 },
\end{aligned}
\label{eq:CordCompodef}
\end{equation}
where the variable $A$ denotes a physical quantity. We note that the correlation is computed with respect to the deviation from the angle-averaged quantities ("$\delta$" in Eq.~\ref{eq:CordCompodef} denotes the deviation).

Let us first take a look at the two quantities $\alpha$ and $\Delta \kappa$. The former is defined as the ratio of the number density of $\bar{\nu}_e$ to $\nu_e$, i.e., $n_{\bar{\nu}_e}/n_{\nu_e}$; the latter denotes the difference of energy-integrated flux factor between $\nu_e$ and $\bar{\nu}_e$. The values of the correlation function of $\alpha$ and $\Delta \kappa$ to $\Delta G_{\rm out}$ are $-0.62$ and $0.73$, respectively. The negative value of the correlation function of $\alpha$ suggests that $\alpha$ tends to be larger (i.e., close to unity), and the positive value for $\Delta \kappa$ indicates that the $\bar{\nu}_e$ angular distributions tend to be more strongly forward-peaked than those for $\nu_e$. Both trends lead to negative $\Delta G_{\rm out}$. The trend is consistent with the correlation found in the bottom left and top right panel of Fig.~\ref{Mol_12M_100ms_130km}.

The origin of angular inhomogeneity can be traced to the convective activity of the fluid. We note that the layer at $100 {\rm km} \lesssim R \lesssim R_{\rm sh}$ is, in general, convectively unstable due to a negative entropy gradient during this phase. As a result, low entropy plumes advect inward from the vicinity of shock wave as a response to rising high-entropy buoyancy bubbles generated at $\sim 100$ km. The low entropy plumes tends to be (adiabatically) compressed to sustain pressure equilibrium with their surroundings. As shown in the middle and bottom right panel, we find that the Type-II crossing appears around the low entropy and high density plumes.

The next question is why the plumes provide a preferable condition for the occurrence of the Type-II crossings. One possible explanation is that high density environment facilitates neutrino-matter interactions, which potentially affects ELN-crossings. However, this effect is common for all species of neutrinos, indicating that it does not fully account for the disparity between $\nu_e$ and $\bar{\nu}_e$. Another key ingredient is the difference of neutrino absorption between $\nu_e$ and $\bar{\nu}_e$. In general, the mass fraction of free neutrons is higher than that of free protons in the post-shock region. This indicates that $\nu_e$ absorptions by free neutrons may overwhelm $\bar{\nu}_e$ absorptions by free protons. Moreover, there is no energy threshold in the $\nu_e+n\rightarrow p+e^-$ channel while there is in the $\bar\nu_e+p\rightarrow n +e^+$ channel. This can make the capture rates sensitive to neutrino or antineutrino energy spectra. While there is no Coulomb correction in the $\bar\nu_e$ capture channel, there is the $\nu_e$ capture channel giving an enhancement, albeit very small ($\sim 1\%$), to the rate in that channel. Likewise, weak magnetism effects further enhance $\nu_e$ capture over $\bar\nu_e$ capture. All of these effects help to accentuate the disparity in the capture rates, all favoring $\nu_e$ capture.

We show radial profiles of angular-averaged $G_{\rm out}$ in the top panel of Fig.~\ref{graph_12M100msAngave}. This illustrates that $G_{\rm out}$ for $\nu_e$ decreases with radius more sharply than for the corresponding quantity for $\bar{\nu}_e$. In the second panel of Fig.~\ref{graph_12M100msAngave}, we show radial profiles of quantities more related to the neutrino absorption rate: the mass density of free neutrons ($\rho_{n}$) and protons ($\rho_{p}$) multiplied by the square of average energy of $\nu_e$ for the former and of $\bar{\nu}_e$ for the latter. 
The  average energy weighting captures the fact that the neutrino-absorption cross section is roughly proportional to the square of neutrino energy, ignoring $Q$-value and threshold issues. 
Although the energy of $\nu_e$ is smaller than that of $\bar{\nu}_e$, we confirm that the difference of mass fraction of protons and neutrons is the dominant factor in the overall absorption rates in the region, overwhelming the threshold difference and weak magnetism effect. Given these considerations, $\alpha$ increases with radius (see the bottom panel of Fig.~\ref{graph_12M100msAngave}), which provides favorable conditions for generating Type-II crossings at $\gtrsim 100$ km.

To see the progenitor dependence, we show the results for the correlation functions of $\alpha$, $\Delta \kappa$, $\rho$, and $s$ to $\Delta G_{\rm out}$ for all CCSN models in Fig.~\ref{graph_Correlation_ELNout_100ms_100km_Rsh_prodepe}. In this figure, we adopt the maximum (minimum) values of the correlation function for the positive (negative) sign in the range of $100 {\rm km} < R < R_{\rm sh}$, where $R_{\rm sh}$ denotes the shock radius. We find that the results are largely progenitor-independent, suggesting that the above arguments are generic trends in CCSNe.

It is worth mentioning that the high mass accretion rate in the early post-bounce phase plays an important role in the above mechanism. By virtue of the high mass accretion rate, the matter density in the post-shock region tends to be higher than that in the later phases. This enhances neutrino-matter interactions. This argument is supported by the fact that the 25 (9) solar mass model has the largest (smallest) peak value of $R_{\rm ELNC}$ among our CCSN models (see the top left panel of Fig.~\ref{graph_Angave_rvsELNC}), meanwhile it has the highest (lowest) mass accretion rate at this time among the models (see Fig.~\ref{graph_Massaccretion_ELN}).

Before closing this subsection, we remark on the weak-magnetism effects on nucleon scatterings that are incorporated in our CCSN simulations. This effect increases (decreases) the scattering cross-sections for $\nu_e$ ($\bar{\nu}_e$), indicating that it boosts the possibility of appearance of Type-II crossings. Although the correction is minor \cite{2002PhRvD..65d3001H}, it may play a non-negligible role on the crossings, since the ELN crossing is very subtle (see also \cite{2021arXiv210602650N}).

\subsubsection{ELN crossings in the layer of PNS convection}\label{subsubsec:PNSconv}

\begin{figure*}
  \begin{minipage}{0.8\hsize}
    \includegraphics[width=\linewidth]{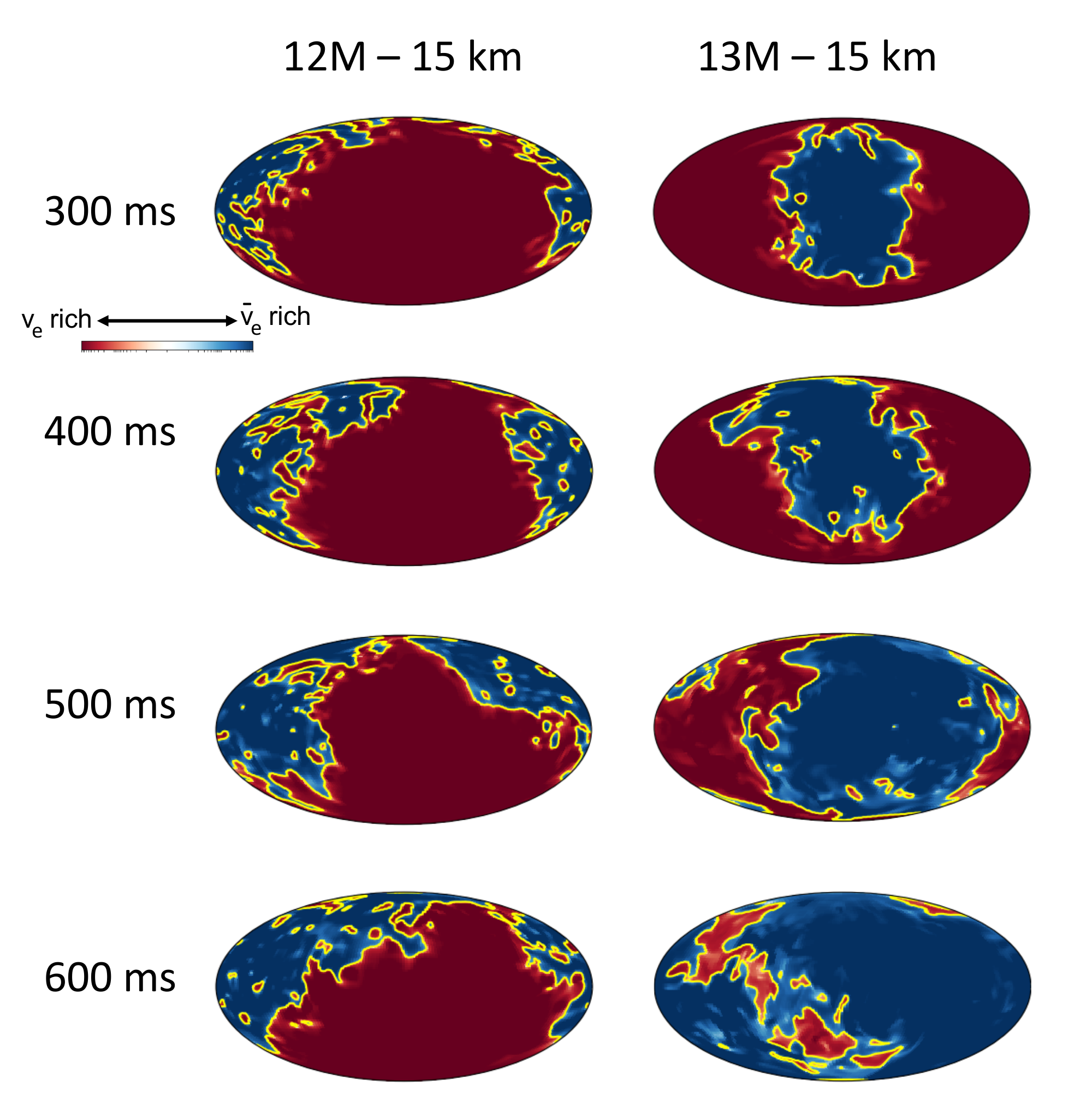}
    \caption{Mollweide projection of ELN number density. We selected two different CCSN models: left and right columns correspond to 12 and 15 solar mass models, respectively. From top to bottom, we show the result at different time; the radius is fixed at $15$ km where PNS convection is in activity. The thick brown and blue denote the $\nu_e$ and $\bar{\nu}_e$ dominant area, respectively. The boundary between the two area is highlighted by yellow thick lines. The boundary layer corresponds to the region where ELN crossings appear. See text for more detail.}
    \label{PNSconv}
  \end{minipage}
\end{figure*}

PNS convection commonly appears in the layer at $10 {\rm km} \lesssim R \lesssim 30 {\rm km}$ at $\gtrsim 200$ ms, regardless of the success or failure of shock revival \cite{2020MNRAS.492.5764N}. We note that there are two important previous works that have investigated fast flavor conversion in the convective layer \cite{2019PhRvD..99j3011D,2020PhRvD.101f3001G}. Our results are essentially consistent with these earlier studies.

Figure~\ref{PNSconv} provides Mollweide projections of ELN number density, i.e., $n_{\nu_e}-n_{\bar{\nu}_e}$, in the PNS convection layer ($15$ km) at multiple time snapshots for 12- and 13 solar mass models. As shown in these maps, there are regions where $\bar{\nu}_e$ dominates over $\nu_e$ (blue-colored area). The emergence of $\bar{\nu}_e$-rich region is one of the important consequences of PNS convection. As discussed in \cite{2020MNRAS.492.5764N}, PNS convection, in general, accelerates the deleptonization; thus $Y_e$ and the chemical potential of $\nu_e$ tend to be smaller (the latter can even become negative in some regions) than those in spherically symmetric models.

The emergence of both $\nu_e$ and $\bar{\nu}_e$ rich regions provides a necessary condition for ELN crossings; the crossings must occur in the vicinity of the boundary of the two regions, i.e., $n_{\nu_e} \sim n_{\bar{\nu}_e}$. In Fig.~\ref{PNSconv}, we highlight them as yellow lines. Because these regions are very optically thick to neutrinos, the neutrino angular distributions are almost isotropic, regardless of flavor. On the other hand, the anisotropy does hinge on neutrino species, and this dependence does provide sufficient conditions for generating the ELN crossings\footnote{It is meaningless to classify the type of ELN crossings appearing in the region of PNS convection. In fact, the number of crossings may be multiple (see also \cite{2021arXiv210101226B}).}. It should be mentioned, however, that the spatial scale of the crossing regions would be very narrow (see an order estimate of the spatial scale in \cite{2020PhRvD.101f3001G}), implying that it is hard to resolve the crossings in our CCSN simulations. This is the main reason why we omit displaying them in Fig.~\ref{graph_Angave_rvsELNC}.

It is interesting to note that the angular profiles of ELN distributions in the 12 solar mass model (the left column in Fig.~\ref{PNSconv}) are less time-dependent than those in the 13 solar mass model (the right one in Fig.~\ref{PNSconv}). This trend is associated with the asymmetric distribution of $Y_e$ and LESA. Indeed, strong dipole lepton number emission is observed in the 12 solar mass model, and the associated dipole direction only becomes significantly time dependent for times $\sim 600$ ms (see Fig. 17 in \cite{2019MNRAS.489.2227V}). In the next subsection (Sec.~\ref{subsubsec:ELNCwithasym}), we discuss how coherent asymmetric neutrino emission triggers another type of ELN crossing outside the layer of PNS convection. Sustaining the coherent profile of lepton number asymmetry may have a substantial influence on nucleosynthesis (see, e.g., \cite{2019MNRAS.488L.114F,2021MNRAS.502.2319F}) and neutrino signals \cite{2021MNRAS.500..696N}. This highlights the potential importance of the impact of fast flavor conversions on nucleosynthesis and the neutrino signals under the long-term and coherent asymmetric neutrino emission. We defer addressing this issue to future work.

\subsubsection{Type-II ELN crossings associated with asymmetric neutrino emission}\label{subsubsec:ELNCwithasym}

\begin{figure*}
  \begin{minipage}{0.8\hsize}
    \includegraphics[width=\linewidth]{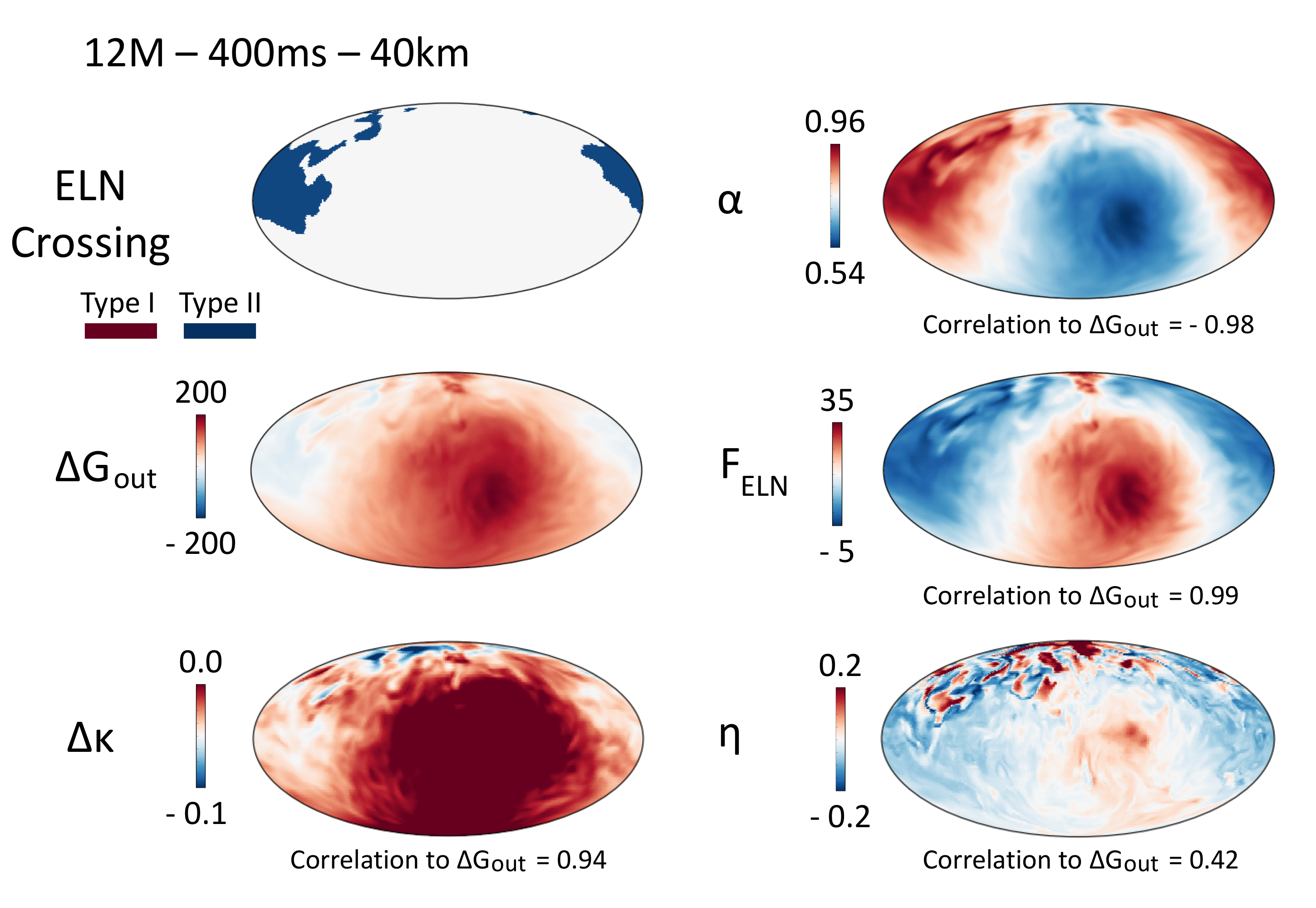}
    \caption{Similar as Fig.~\ref{Mol_12M_100ms_130km} but for different radius ($40$ km) and time ($400$ ms). In this figure, we show the Mollweide projection for $F_{\rm ELN}$ and $\eta$ (instead of $\rho$ and $s$ in Fig.~\ref{Mol_12M_100ms_130km}). $F_{\rm ELN}$ and $\eta$ denote the energy-integrated number flux of ELN and the degeneracy of $\nu_e$, respectively. The result of correlation function to $G_{\rm out}$ is also inserted at each panel.}
    \label{Mol_12M_400ms_40km}
  \end{minipage}
\end{figure*}

Anti-correlated asymmetric neutrino emission between $\nu_e$ and $\bar{\nu}_e$ provides favorable conditions for generating ELN crossings (see also \cite{2019ApJ...886..139N}). This anti-correlation is a general trend if the asymmetry is driven and sustained by the dynamics in the vicinity of the PNS. This is attributed to the fact that the neutrino emission for these species is closely associated with $Y_e$ distributions. As demonstrated in \cite{2019ApJ...880L..28N}, the PNS kick breaks the symmetry of the matter distribution around the PNS. In turn, this helps generate the large-scale asymmetry in $Y_e$ distributions. Asymmetry of $Y_e$ is also commonly observed with LESA regardless of the detailed generation mechanism \cite{2014ApJ...792...96T,2018ApJ...865...81O,2019MNRAS.487.1178P,2019MNRAS.489.2227V,2019ApJ...881...36G}. We note that the LESA is observed in our 3D CCSN models (see also \cite{2019MNRAS.489.2227V}). LESA potentially plays an important role in generating ELN crossings in these models.

We find that Type-II crossings observed in the region of $\gtrsim 40$ km at $\gtrsim 300$ ms (see also in Fig.~\ref{graph_Angave_rvsELNC}) can be attributed to asymmetric neutrino emission. To see the trend more clearly, we provide Mollweide projection of some important physical quantities for the 12 solar mass model in Fig.~\ref{Mol_12M_400ms_40km}. The selected time and radius in the figure are $400$ ms and $40$ km, respectively. The new variables $F_{\rm ELN}$ and $\eta$ displayed in the middle and bottom right panels of Fig.~\ref{Mol_12M_400ms_40km} denote the energy-integrated number flux of ELN and the degeneracy parameter for $\nu_e$, respectively. The latter is defined as a chemical potential of $\nu_e$ divided by matter temperature.

In Fig.~\ref{Mol_12M_400ms_40km}, we find some evidences that the Type-II ELN crossings are associated with LESA: $\alpha$ tends to be close to 1: $F_{\rm ELN}$ becomes smaller in the region of the crossings: $\Delta \kappa$ tends to be negative. All these trends suggests that the crossing appears in the direction opposite to that of the LESA dipole. To quantify the correlation, we compute the correlation function by following Eq.~\ref{eq:Cordef}. The values of each correlation function are $> 0.9$, indicating that they are strongly correlated with $G_{\rm out}$.

It is worth mentioning that the correlation of local matter quantities to $G_{\rm out}$ is much weaker than for the above three quantities. The highest correlated quantity among them is $\eta$ (see the bottom right panel of Fig.~\ref{Mol_12M_400ms_40km}); the correlation function is $\sim 0.4$. This relatively weak correlation suggests that the ELN crossing at this radius is not primarily driven by local phenomena but rather by a global one. This is consistent with our interpretation that the ELN crossings are mainly driven by LESA. Indeed, the LESA is develops in the inner region ($\lesssim 20$ km, see also \cite{2019ApJ...881...36G,2019MNRAS.487.1178P}).

Although asymmetric neutrino emission commonly appears in our CCSN models, the same type of crossing is not observed in failed explosion models. This is due primarily to the fact that $\alpha$ in non-exploding models tends to be smaller than in the exploding ones, which suppresses the occurrence of ELN crossings. The high $\nu_e$ population stems mostly from the continuous supply of high-$Y_e$ matter from accretion. The fresh high-$Y_e$ accreted matter is efficiently deleptonized by electron captures on free protons in the post-shock region and, as a consequence, $\nu_e$ emission becomes higher than $\bar{\nu}_e$ emission.

\subsubsection{Type-I ELN crossings in exploding models}\label{subsubsec:TypeIexplo}

\begin{figure*}
  \begin{minipage}{0.8\hsize}
    \includegraphics[width=\linewidth]{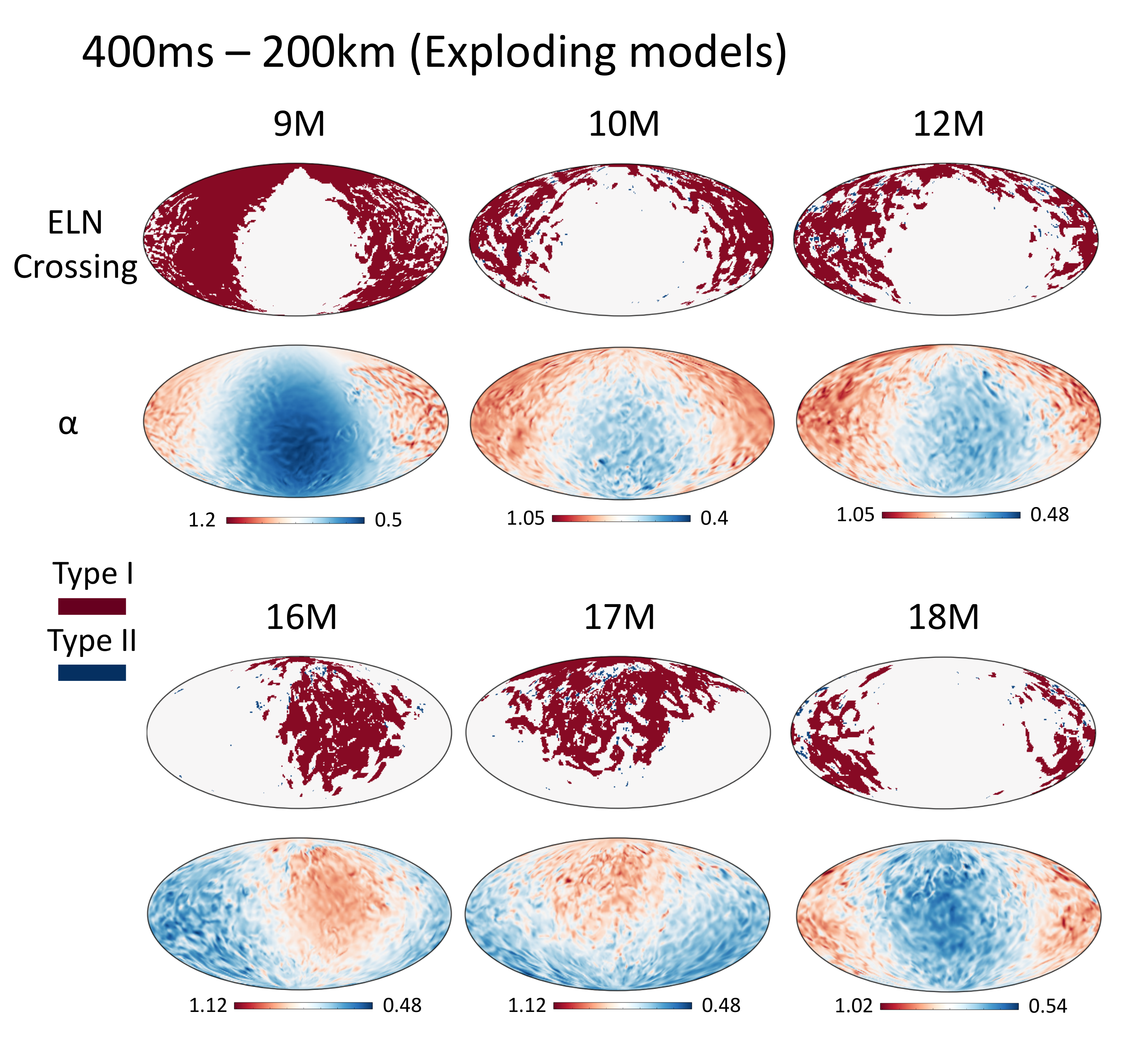}
    \caption{Mollweide projection of ELN crossings and $\alpha$ at the radius of $200$ km in the time snapshot of $400$ ms after bounce. We show the result of exploding models that the development of explosion has been already matured by that time.}
    \label{Mol_Exprodepe_400ms_200km}
  \end{minipage}
\end{figure*}

\begin{figure}
  \begin{minipage}{1.0\hsize}
    \includegraphics[width=\linewidth]{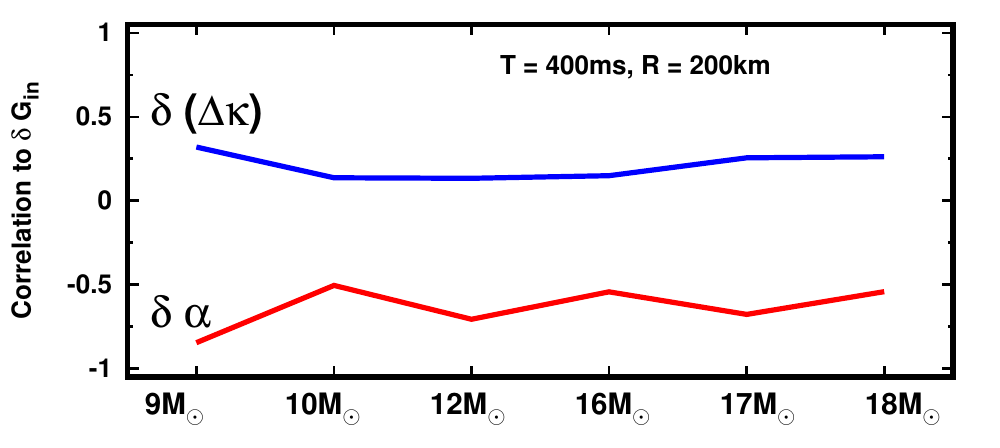}
    \caption{Correlation function of $\Delta \kappa$ (blue) and $\alpha$ (red) to $G_{\rm in}$. They are computed at the radius of $200$ km and in the time snapshot of $400$ ms.}
    \label{graph_Correlation_ELNin_400ms_200km_prodepe}
  \end{minipage}
\end{figure}

\begin{figure*}
  \begin{minipage}{1.0\hsize}
    \includegraphics[width=\linewidth]{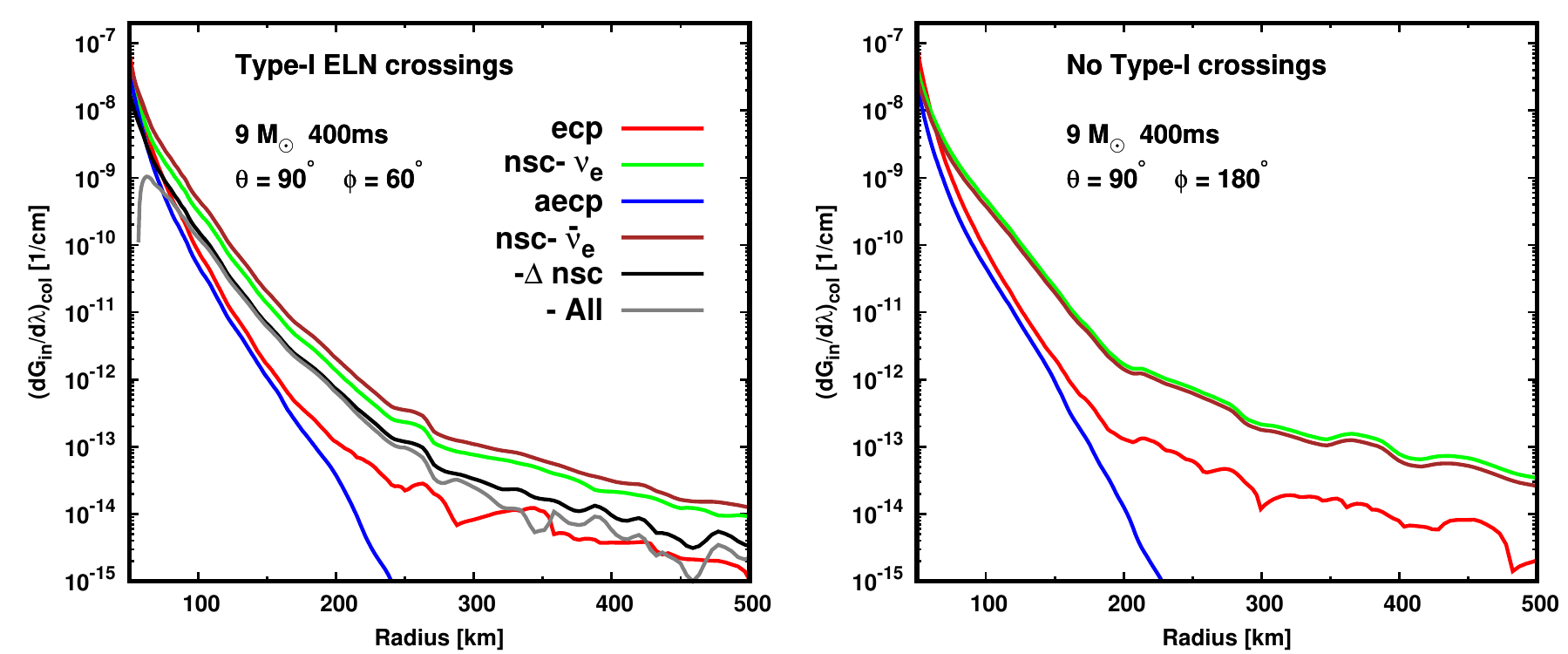}
    \caption{Radial profiles of energy-integrated collision term for incoming neutrinos ($dG_{\rm in}/d\lambda$), in which the unit is measured with ${\rm cm}^{-1}$. We show the result of 9 solar mass model (exploding model). In both panels, we use the same time snapshot ($400$ ms) but different radial rays. In the radial ray with $\theta=90^{\circ}$ and $\phi=60^{\circ}$ (left panel), we find Type-I crossing in the region of $\gtrsim 50$ km, whereas the crossings are not observed in the radial ray with $\theta=90^{\circ}$ and $\phi=180^{\circ}$ (right panel). Color distinguishes neutrino-matter interaction; red, green, blue, and brown represent electron capture by free protons (ecp), nucleon scattering of $\nu_e$ (nsc-$\nu_e$), positron capture by free neutrons (aecp), and nucleon scattering of $\bar{\nu}_e$ (nsc-$\bar{\nu}_e$), respectively. The inverse reactions (neutrino absorptions and out-scatterings) are also taken into account. Black line represents ``$-\Delta {\rm nsc}$'', which portrays the difference of nucleon scatterings between $\bar{\nu}_e$ and $\nu_e$. It should be noted that the positive sign of ``$-\Delta {\rm nsc}$'' indicates that nsc-$\bar{\nu}_e$ dominates over nsc-$\nu_e$. The gray line represents the collision term with respect to ELN, and we change the sign of the rate in the plot. Similar as ``$-\Delta {\rm nsc}$'', the positive sign of ``$- {\rm All}$'' indicates that $\bar{\nu}_e$ increases more rapidly than $\nu_e$ during the propagation of neutrinos (inward direction). In the right panel, both ``$-\Delta {\rm nsc}$'' and ``$- {\rm All}$'' are negative; hence, these lines are not displayed in the panel.}
    \label{graph_reaccompare_s9_400ms}
  \end{minipage}
\end{figure*}

\begin{figure}
  \begin{minipage}{1.0\hsize}
    \includegraphics[width=\linewidth]{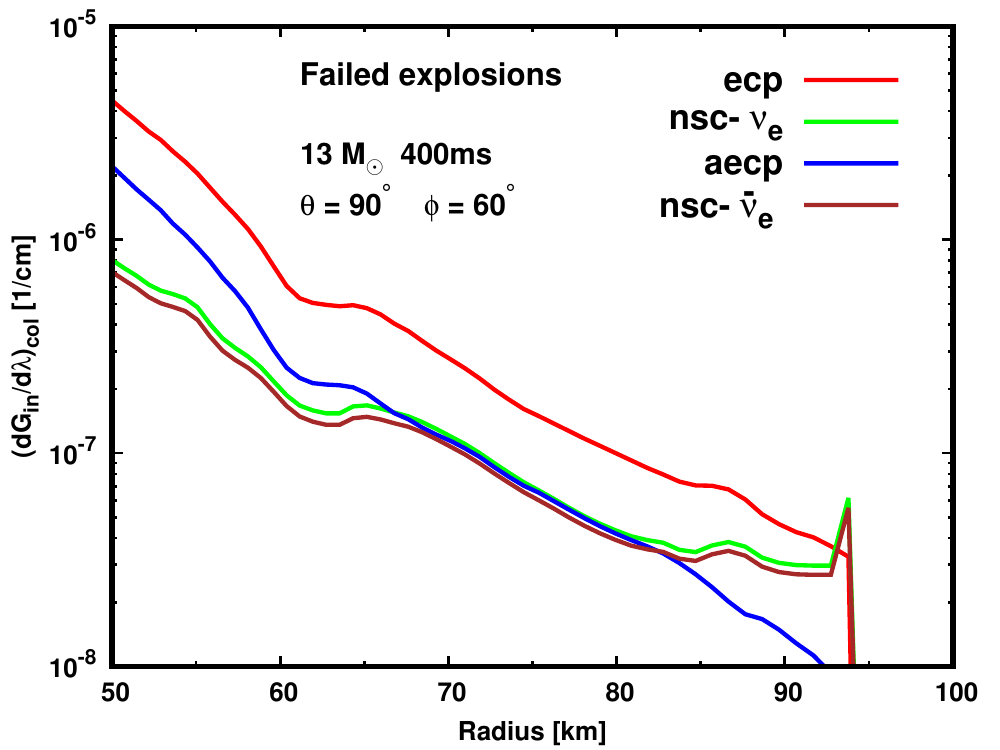}
    \caption{Same as Fig.~\ref{graph_reaccompare_s9_400ms} but for 13 solar mass model (no explosion model). We show the result along a radial ray with $\theta=90^{\circ}$ and $\phi=60^{\circ}$. We do not observe Type-I ELN crossings at this phase in the model.}
    \label{graph_reaccompare_s13_400ms}
  \end{minipage}
\end{figure}

Finally, we discuss the origin of Type-I ELN crossings appearing only in exploding models. Similar to the previous analyses, we plot Mollweide projection of ELN crossings and $\alpha$ at $200$ km for the time snapshot at $400$ ms after bounce in Fig.~\ref{Mol_Exprodepe_400ms_200km}. We selected $9, 10, 12, 16, 17,$ and $18$ solar mass models for display in the figure, since for these models the development of an explosion is well along by $400$ ms after bounce.

The figure illustrates that Type-I ELN crossings appear in the region of $\alpha \sim 1$ for all the progenitors considered. To see the correlation quantitatively we compute the appropriate correlation function following Eq.~\ref{eq:Cordef}. It should be noted, however, that Type-I ELN crossings are mainly characterized by incoming neutrinos. Hence we replace $G_{\rm out}$ with $G_{\rm in}$ in the computation of the correlation function. As shown in Fig.~\ref{graph_Correlation_ELNin_400ms_200km_prodepe}, the ELN crossing is mainly associated with $\alpha$, whereas $\Delta \kappa$ plays a subdominant role in the crossings. We note that the negative correlation of $\alpha$ with $G_{\rm in}$ implies that $\alpha$ tends to be high in the region of the small $G_{\rm in}$, which is consistent with what we observed in Fig.~\ref{Mol_Exprodepe_400ms_200km}. It should also be mentioned that asymmetric neutrino emission facilitates the generation of the Type-I crossings in a manner similar to the Type-II case discussed in Sec.~\ref{subsubsec:ELNCwithasym}. Indeed, the crossings tend to appear in the region of higher $\bar{\nu}_e$ emission\footnote{There is a caveat, however. If the $\bar{\nu}_e$ emission is much higher than $\nu_e$, $\bar{\nu}_e$ becomes dominant in all directions, which erases Type-I crossings.}.

Ray-tracing neutrino transport provides key information in understanding the origin of the Type-I ELN crossings. In Fig.~\ref{graph_reaccompare_s9_400ms}, we show the radial profile of the energy-integrated collision term for each reaction with respect to the neutrino flight direction with $\mu=-1$. In the figure, we display the result for the 9 solar mass (exploding) model at $400$ ms. The left panel corresponds to the collision term along a radial ray of $\theta=90^{\circ}$ and $\phi=60^{\circ}$ (Type-I crossings appear at $\gtrsim 50$ km), whereas the right panel shows the result along another radial ray with no ELN crossings ($\theta=90^{\circ}$ and $\phi=180^{\circ}$).

As shown in both panels, the dominant neutrino-matter interaction at $\gtrsim 100$ km is nucleon scattering (nsc) for both $\nu_e$ and $\bar{\nu}_e$. The next dominant one is electron-capture by free protons and its inverse reaction (ecp). Positron-capture by free neutrons (aecp) is subdominant at large radius ($\gtrsim 200$ km) but is non-negligible in the inner region. It should be noted that we omit the contribution of coherent scatterings by heavy nuclei in this figure, since it plays a negligible role due to the small heavy nucleus mass fraction. One of the qualitative differences between the neutrino-matter interactions along the two radial rays is that the ``nsc'' of $\bar{\nu}_e$ overwhelms that of $\nu_e$ in the left panel, whereas the opposite trend is observed in the right one. In other words, the ELN in the direction of $\mu=-1$ tends to be negative because of ``nsc'' in the left panel, and that potentially plays a role in driving the Type-I crossings. To see the impact more quantitatively, we further compute the difference of ``nsc'' between $\nu_e$ and $\bar{\nu}_e$ ($\Delta {\rm nsc}$). This result is also displayed in the same panel (see the black line in the left panel of Fig.~\ref{graph_reaccompare_s9_400ms}\footnote{We note that the sign of $\Delta {\rm nsc}$ is negative; hence, we change the sign in the figure.}). As shown in the figure, ``$-\Delta {\rm nsc}$'' is higher than ``ecp'' at $\gtrsim 80$ km. This suggests that the ELN in the direction of $\mu=-1$ tends to be negative, propagating inward, leading to Type-I crossings.

In the right panel, on the other hand, we observe that $\Delta {\rm nsc}$ is positive, i.e., ``nsc'' increases the ELN in the direction of $\mu=-1$, which suppresses the appearance of Type-I crossings. The positive sign of $\Delta {\rm nsc}$ suggests that $\bar{\nu}_e$ in $\mu=1$ direction is much smaller than $\nu_e$. Although the cross section of the $\bar{\nu}_e$ is higher than $\nu_e$ (since the average energy of the former is higher than the latter), the low population of $\bar{\nu}_e$ reduces the total frequency of scatterings. Consequently, ``nsc'' of the $\nu_e$ becomes higher than that of the $\bar{\nu}_e$ along this ray. This picture is consistent with the trend observed in Fig.~\ref{Mol_Exprodepe_400ms_200km}: $\alpha$ is much smaller than unity in the area with no Type-I crossings.

Figure~\ref{graph_reaccompare_s9_400ms} also provides a key information to understand the disappearance of Type-I ELN crossings at inner regions ($\lesssim 50$ km) (see Sec.~\ref{subsec:overall} and Fig.~\ref{graph_Angave_rvsELNC}). As shown in the left panel of Fig.~\ref{graph_reaccompare_s9_400ms}, "esc" gradually dominates over ``nsc'' with decreasing radius. As a result, ELN in $\mu=-1$ becomes positive, i.e., Type-I ELN crossings disappear. Although the reaction rate of ``aecp'' rate also increases with decreasing radius, it does not overwhelm the ``ecp''. The sharper increase of those reactions is mainly due to finite-temperature effects. Since the temperature dependence of both ``ecp'' and ``aecp'' reactions is stronger than that of ``nsc'', these reactions become dominant at the inner region.

The temperature dependence of ``ecp'' is also responsible for another trend: the same type (Type-I) ELN crossing does not appear in failed explosion models. In Fig.~\ref{graph_reaccompare_s13_400ms}, we show the same quantities as in Fig.~\ref{graph_reaccompare_s9_400ms} but for the 13 solar mass model (failed explosion model). As clearly seen in the plot, ``ecp'' dominates over other reactions, which hampers generating Type-I ELN crossings. Since the post-shock flow in non-exploding models has more compact structures than exploding models, the matter temperature tends to be higher in the entire post-shock region. This helps facilitate ``ecp'' in becoming the dominant neutrino-matter reaction. It should also be noted that a previous study \cite{2021PhRvD.103f3033A} suggested that no ELN crossings arising in failed CCSN models stems mainly from isotropic distributions of neutrinos. Our study suggests that this is a misinterpretation. In fact, the angular distribution of neutrinos at $\gtrsim 50$ km is strongly anisotropic at this phase regardless of neutrino species\footnote{The flux factor of energy-integrated neutrinos in the region of $50 {\rm km} < R < 100 {\rm km}$ is roughly from $0.7$ to $0.9$.}. It is attributed to the fact that the matter density in the post-shock flow is low compared to that in the early phase ($\sim 100$ ms) owing to the low mass accretion rates. This suggests that the neutrino-matter interaction has less influence on the out-going neutrinos. The key ingredient in accounting for this trend is not the degree of isotropy of the neutrino angular distributions but the dominance of ``ecp'' over ``nsc'' in the post-shock region, that plays a crucial role in determining ELN for incoming neutrinos.

The above argument also leads to a general conclusion. The Type-I ELN crossings tend to appear regions with conditions where $\alpha$ is close to unity and where the matter temperature is cold. These findings may be useful in using non-detailed analyses to find places which would be favorable for the occurrence of fast flavor conversions.

\section{Summary}\label{sec:summary}
In this paper, we have investigated the appearance of ELN crossings as a good indicator of the occurrence of fast flavor conversions. We have analyzed along these lines more than a dozen 3D CCSN models. Although the neutrino data from our CCSN models are limited to energy-dependent zeroth and first angular moments, we have developed a new ELN crossing search method, essentially a hybrid method of two-moment and ray-tracing approaches (see \cite{2021arXiv210602650N} for more detail). This has enabled us to find ELN crossings from limited angular information. Our method allows us to utilize the time-, space-, energy-, angular-, and species-dependent structures of neutrino radiation fields without solving the full Boltzmann neutrino transport equation. It provides quantitative information on the physical origin of each type of ELN crossing. By virtue of its high-fidelity, the method reveals the generation mechanism of each type of ELN crossing.

Below, we summarize our findings:
\begin{enumerate}
\item We find ELN crossings in the post-shock region for all models at $\gtrsim 100$ ms after bounce. We summarize the trends in Fig.~\ref{PhaseDiagram}.
\item Type-II ELN crossings (see Fig.~\ref{graphType} for the classification of ELN crossings) in the early post-bounce phase ($\sim 100$ ms) commonly appear right behind the shock wave (see the top left panel of Fig.~\ref{graph_Angave_rvsELNC}). This is mainly due to the disparity between absorptions of $\nu_e$ and $\bar{\nu}_e$ neutrinos. The high mass accretion rate in this phase also provides conditions favorable for the generation of this type of crossing.
\item The ELN crossings are ubiquitous in the PNS convection layer. This result is consistent with the findings of previous studies \cite{2020PhRvD.101b3018D,2020PhRvD.101f3001G}. We also find that the spatial pattern of ELN crossing distributions is less time-dependent if coherent asymmetric neutrino emission occurs (see Fig.~\ref{PNSconv}). This motivates us to study in future work the impact of fast flavor conversions on nucleosynthesis and neutrino signals.
\item Asymmetric neutrino emission provides a favorable condition for the generation of ELN crossings, consistent with the result of \cite{2019ApJ...886..139N}. Our correlation study (see Fig.~\ref{Mol_12M_400ms_40km}) suggests that the LESA is mainly responsible for Type-II crossings appearing at $\gtrsim 40$ km in our exploding models. On the other hand, this type of crossing does not appear in failed explosion models, since $\nu_e$ emission tends to be much stronger than that for $\bar{\nu}_e$ due to the continuous supply of high-$Y_e$ matter from accretion.
\item Type-I ELN crossings are observed in exploding models. They are driven mainly by nucleon scattering. We find that they tend to appear in regions of $\alpha \sim 1$ and low matter temperature. In Sec.~\ref{subsubsec:TypeIexplo}, we provided a rationale for this conclusion based on the result of ray-tracing neutrino transport (see Fig.~\ref{graph_reaccompare_s9_400ms}).
\item Type-I crossings, that appear in exploding models, are not observed in non-exploding models. This is mainly due to the dominance of electron captures by free protons over nucleon scatterings. Since the post-shock flows in non-exploding models are more compact structures, the matter temperature tends to be higher in these models and, in turn, that higher temperature facilitates electron captures by free protons and that helps erase Type-I crossings.
\end{enumerate}

Finally, we need to mention important caveats. Our overall conclusions still suffer from many uncertainties, including uncertainties in the CCSN models, input physics, and our ELN crossing search method. It will be important for us to reduce each of these uncertainties to achieve a comprehensive understanding of fast flavor conversions in CCSNe. Nevertheless, the results of our present study suggest that fast flavor oscillation commonly occurs in the post-shock region of CCSNe. This highlights the need for more detailed investigation of the non-linear evolution and feedbacks in CCSN dynamics, nucleosynthesis, and neutrino signals. Although there are many technical challenges ahead, it is clear that the study of collective neutrino oscillations is in a state of rapid development. Future efforts may provide a more robust picture of the CCSN explosion mechanism or may change our basic understanding of CCSN dynamics.

\section{Acknowledgments}
We are grateful to Sherwood Richers for useful comments and discussions. We are also grateful for ongoing contributions to our CCSN project by David Vartanyan, David Radice, Josh Dolence and Aaron Skinner, Evan O'Connor regarding the equation of state, Gabriel Mart\'inez-Pinedo concerning electron capture on heavy nuclei, Tug Sukhbold and Stan Woosley for providing details concerning the initial models, and Todd Thompson regarding inelastic scattering. We acknowledge support from the U.S. Department of Energy Office of Science and the Office of Advanced Scientific Computing Research via the Scientific Discovery through Advanced Computing (SciDAC4) program and Grant DE-SC0018297 (subaward 00009650). In addition, we gratefully acknowledge support from the U.S. NSF under Grants AST-1714267 and PHY-1804048 (the latter via the Max-Planck/Princeton Center (MPPC) for Plasma Physics). An award of computer time was provided by the INCITE program. That research used resources of the Argonne Leadership Computing Facility, which is a DOE Office of Science User Facility supported under Contract DE-AC02-06CH11357. In addition, this overall research project is part of the Blue Waters sustained-petascale computing project, which is supported by the National Science Foundation (awards OCI-0725070 and ACI-1238993) and the state of Illinois. Blue Waters is a joint effort of the University of Illinois at Urbana-Champaign and its National Center for Supercomputing Applications. This general project is also part of the ``Three-Dimensional Simulations of Core-Collapse Supernovae" PRAC allocation support by the National Science Foundation (under award \#OAC-1809073). Moreover, access under the local award \#TG-AST170045 to the resource Stampede2 in the Extreme Science and Engineering Discovery Environment (XSEDE), which is supported by National Science Foundation grant number ACI-1548562, was crucial to the completion of this work. Finally, the authors employed computational resources provided by the TIGRESS high performance computer center at Princeton University, which is jointly supported by the Princeton Institute for Computational Science and Engineering (PICSciE) and the Princeton University Office of Information Technology, and acknowledge our continuing allocation at the National Energy Research Scientific Computing Center (NERSC), which is supported by the Office of Science of the US Department of Energy (DOE) under contract DE-AC03-76SF00098. L.J. was supported by NASA through the NASA Hubble Fellowship Grant Number HST-HF2-51461.001-A awarded by the Space Telescope Science Institute, which is operated by the Association of Universities for Research in Astronomy, Incorporated, under NASA contract NAS5-26555. GMF acknowledges support from NSF Grant No. PHY-1914242, from the Department of Energy Scientific Discovery through Advanced Computing (SciDAC-4) grant register No. SN60152 (award number de-sc0018297), and from the NSF N3AS Hub Grant No. PHY-1630782 and Heising-Simons Foundation Grant No. 2017-22, and the N3AS Physics Frontier Center NSF PHY-2020275.
\bibliography{bibfile}


\end{document}